UDC 372.8

# Rapid Mental Computation System as a Tool for Algorithmic Thinking of Elementary School Students Development

[1] Rushan Ziatdinov
[2] Sajid Musa

[1-2] Fatih University, Turkey
34500 Büyükçekmece, Istanbul
[1] PhD (Mathematical Modelling), Assistant Professor
E-mail: rushanziatdinov@gmail.com
[2] Research Student
E-mail: sajid_musa2003@yahoo.com

**Abstract.** In this paper, we describe the possibilities of using a rapid mental computation system in elementary education. The system consists of a number of readily memorized operations that allow one to perform arithmetic computations very quickly. These operations are actually simple algorithms which can develop or improve the algorithmic thinking of pupils. Using a rapid mental computation system allows forming the basis for the study of computer science in secondary school.

**Keywords:** algorithmic thinking; rapid mental computation; Trachtenberg system; informatics education; mathematics education.

**1. Introduction**

It is well-known that an *algorithm* is a fundamental concept in preparing informatics teachers. Learning algorithmic thinking can start in the early years of a child's life and must be oriented around the thinking ability of young children. Despite this fact, the development of an algorithmic style of thinking encounters many problems such as, for example, in South Korea where elementary school students have a weakness in algorithm and modelling (Cha et al., [6]). In the Czech Republic, where the user approach by students has been increasing, still has the problem that the algorithmic approach is almost unknown to them (Milkova, [17]).

Algorithmic thinking is regarded as a representation of the sequence of actions, along with imaginative and logical thinking that defines the intellectual power of man, i.e., his creativity. Planning skills, the habit of accurate and complete description of their actions, can help students to develop algorithms for solving problems of different origins. Algorithmic thinking is a necessary part of the scientific world view. At the same time, it also includes some general thinking skills that are useful in a broader context, these include, for example, splitting a task into subtasks. For student teaching algorithmic thinking, only the ability to perform arithmetic operations on integers is needed. Knowledge can be the active use of games, theatrical problems.

Futschek et al. [10] present in a learning scenario *Tim the Train* for primary school children that involves tangible objects and allows a variety of interesting tasks designed to assist in the learning of basic concepts of algorithmic thinking. Kolczyk [9] discusses some useful examples of tools and techniques which future informatics teachers should be familiar with. An interesting work by Knuth [16] deals with philosophical questions connected with the actual role of the notion of an algorithm in mathematical sciences, as well as its understanding by computer scientists. Cooper et al. [8] have introduced *Alice*, a 3-dimensional animation tool that provides a learning environment which may be helpful in developing algorithmic thinking. Hubálovský et al. [20] have considered algorithmic learning as an example of a physical problem and used step by step modelling and several approaches as a solution to this. This approach, as well as our case, demonstrates possible interdisciplinary learning, which is considered to be a very important part of future teachers' education. The Australian Informatics Competition (AIC), which has the unique feature of three-stage tasks that invite algorithmic thinking by posing similar problems of increasing size, has been described by Burton [5].

Schwank [19] noted that differences in students' construction and analysis of mathematical algorithms may be explained by the differences between predicative and functional cognitive structures. Moreover, recently Futschek [11] has shown that algorithmic thinking is a key ability in informatics that can be developed independently from learning programming, and this was great motivation for us to investigate if there are any other effective methods except programming that can improve algorithmic thinking.

**2. Existing rapid mental computation systems**

Aside from the Trachtenberg system, several other techniques of rapid mental computation exist. These are Vedic mathematics (Agrawala, [1]) and the Mental Abacus (Stigler, [20]) etc. However, we will not deal with these much as our main subject is the Trachtenberg system. We will simply give a brief and concise idea of the other two techniques mentioned. Vedic mathematics originated in India. Inspired by the Vedas, or sacred Hindu texts, Bharati Krishna Tirthaji (1881-1960), a saint-yogi, researched the ancient Indian scriptures and drew forth 16 sutras, or word-formulae, thus creating Vedic mathematics. It is designed to utilize the right side of the brain; hence making math more assessable to larger audiences. Like the Trachtenberg system, Vedic mathematics is a mental tool for calculation that encourages the development and use of intuition and innovation, while allowing the student a lot of flexibility, fun and satisfaction. With such techniques, the students do not just learn, they also enjoy themselves as they play with numbers and their mental calculation abilities.

Calculating numbers, such as 998 x 997 in less than five seconds flat, is one of the best examples of Vedic mathematics. It also helps in solving many mathematical problems in the branches of arithmetic, algebra, calculus and even geometry. With its extremely fast way of calculation, Vedic mathematics soon became known as "High Speed Vedic Mathematics". It has paved the way for the success of the students in many mathematical competitions and examinations.

With regard to the Mental Abacus, as we all might know, the abacus originated in China. Even before the 14th century the Chinese were using the abacus as their computing device in all trade and business dealings they were associated with. Six centuries have passed and the abacus, which is an ancient calculating device, is still being used in China and Japan. Nowadays, there are many skilled abacus operators and they visualize a mental image of the abacus, performing rapid mental calculations by manipulating the beads on their "mental abacus". Eleven-year-old Chinese children at three levels of abacus expertise were observed performing both abacus and mental addition. Response times and errors were examined as a function of problem type and mode of computation. The results indicate that abacus training has both quantitative and qualitative effects on children's mental calculation skills. However, the "mental abacus" is used by experts (Stigler, [20]) and before one tries the mental abacus, one must be an abacus user with great experience.

The Mathematics of OZ: Mental Gymnastics from Beyond the Edge was created and designed by Clifford Pickover, Dorothy and Dr. Oz (Pickover, [18]). The tests devised by Dr. Oz to assess human intelligence are used to enhance the brain of even the most avid puzzle fan. The Mathematics of Oz focuses on a variety of topics: geometry and mazes, sequences, series, sets, arrangements, probability and misdirection, number theory, arithmetic, and even several problems dealing with the physical world. As many of children struggle in mathematics nowadays, Speed Math for Children was created especially with them in mind. Tricks for understanding fractions and decimals are given, furthermore, checking answers right after every calculation is also taught. A new and revised edition of Speed Math for Children features new chapters on memorizing numbers and general information, calculating statistics and compound interest, square roots, logarithms and easy trig calculations. Created so that anyone can understand, this book teaches simple strategies that will boost calculation skills. Making math easy and fun is possible with speed mathematics and this book fits those who enjoy working with figures and even those who are terrified of numbers.

Chuk Lotta has presented the 250 ten-minute quizzes he developed to help boost his students' mental math skills and their scores on standardized tests. Topics covered include addition, subtraction, multiplication, division, numeration, patterns, per cents, ratios, rounding, prime numbers, geometry and much more.

The amazing mathematical system discussed by Harry Lorayne makes it possible to become a math whiz and can have you solving problems in addition, subtraction, division and multiplication as fast as any calculator and with an even greater degree of precision. Based on the same idea used in the Asian abacus, it requires no more equipment than a pencil and paper.

Mathemagics (Benjamin et al., [2]) shows how to add, subtract, multiply and divide faster in your head than with a calculator, let alone using pencil and paper. With Mathemagics these tricks are so easy to learn that they make calculating actually enjoyable.

Rapid Math contains a few of time-saving tips and tricks for performing common math calculations. It contains sample problems for each trick, leading the reader through step by step. It also has sections on "Mathematical Curiosities" and "Parlor Tricks" created by Edward H. Julius (Julius, [15]).

**3. The basics of the Trachtenberg rapid mental computation system**

In this section we will be seeing an alternative way of multiplication without using any of the multiplication table rules. The system uses some of the basic operations, such as addition and subtraction, and associates them with some unique rules which are actually simple algorithms containing *if-then-else-else* if as conditional statements.

**Rule 3.1 (Multiplication by 11)** The last number of the given digit is put down as the right-hand figure of the answer. Each consecutive digit of the given number or the multiplicand is added to its neighbour at the right. The first digit of the multiplicand becomes the left-hand number of the answer. Then we get the final answer.

**Example 3.1**

|     | 1   | 2   | 3   |
| --- | --- | --- | --- |
| 1+0 | 1+2 | 2+3 | 3+0 |
| 1   | 3   | 5   | 3   |

**Example 3.2**

|       | 4        | 9        | 7   |
| ----- | -------- | -------- | --- |
| 0+4   | 4+9=(1)3 | 9+7=(1)6 | 7+0 |
| 4+(1) | 3+(1)    | 6        | 7   |
| 5     | 4        | 6        | 7   |

**Rule 3.2 (Multiplication by 12)** Double every digit in turn and add its neighbour.

**Example 3.3**

|       | 4          | 9          | 7          |
| ----- | ---------- | ---------- | ---------- |
| 0+4   | 4×2+9=(1)7 | 9×2+7=(2)5 | 7×2+0=(1)4 |
| 4+(1) | 7+(2)      | 5+(1)      | 4          |
| 5     | 9          | 6          | 4          |

Before we proceed, there are some terms that we will be using a lot, such as the left-hand figure, neighbour and right-hand figure. Left-hand figure means the top left digit of the given example. Neighbour indicates the next left number of the digit which we are using. While the right-hand figure is the top right digit of the given number.

As we continue to grasp the rules, we must keep in mind the difference between the even and odd numbers, because we will be using even numbers: 0,2,4,6,8 and the odd numbers: 1,3,5,7,9 a lot in the succeeding rules. Based on these numbers, we will be encountering different rules based on their group.

Another important rule we have to keep in mind is if we take half of the digits, especially for the odd numbers, we throw away the 1 over 2. For example half of 3 is 1. It is really 1 and 1 over 2, but as we apply the rule, we throw away 1 over 2. The same thing goes for 5 which is 2, 7 which is 3 and 9 which is 4. But for even numbers like 2, 4 and the others, there is no need for such a rule.

**Rule 3.3 (Multiplication by 6)** In every digit add the half of its neighbour and add another 5 if the digit is odd.

**Example 3.4**

|        | 4      | 9           | 7            |
|--------|--------|-------------|--------------|
| 0+2    | 4+4=8  | 9+3+5=(1)7  | 7+0+5=(1)2   |
| 2      | 8+(1)  | 7+(1)       | 2            |
| 2      | 9      | 8           | 2            |

**Rule 3.4 (Multiplication by 7)** Double the digit and add half of its neighbour, and add another 5 if the digit is odd.

**Example 3.5**

|              | 4             | 9               | 7                |
|--------------|---------------|-----------------|------------------|
| 0×2+2=2      | 4×2+4=(1)2    | 9×2+3+5=(2)6    | 7×2+0+5=(1)9     |
| 2+(1)        | 2+(2)         | 6+(1)           | 9                |
| 3            | 4             | 7               | 9                |

**Rule 3.5 (Multiplication by 5)** Take the half of its neighbour and add another 5 if the number is odd.

**Example 3.6**

|     | 4     | 9     | 7     |
|-----|-------|-------|-------|
| 2+0 | 4+0=4 | 3+5=8 | 0+5=5 |
| 2   | 4     | 8     | 5     |
| 2   | 4     | 8     | 5     |

As we notice we only add another 5 for those numbers which are odd, but for the even numbers they remain as usual and we will not add any 5 to it.

**Rule 3.6 (Multiplication by 9)** Subtract the right-hand figure of the given number by ten. Taking the remaining numbers up to the last one, subtract it from 9 and add the neighbour. Finally, when you are left with zero in front of the given number, subtract 1 from the neighbour and that serves as the final digit of the answer.

**Example 3.7**

|       | 4          | 9       | 7      |
|-------|------------|---------|--------|
| 4-1=3 | 9-4+9=(1)4 | 9-9+7=7 | 10-7=3 |
| 3+(1) | 4          | 7       | 3      |
| 4     | 4          | 7       | 3      |

**Rule 3.7 (Multiplication by 8)** Subtract the right-hand figure from ten and double it. Taking the remaining numbers up to the last 1, subtract it from 9 and double it, when you get the result add the neighbour to it. For the left-hand figure, subtract 2 from it and then we get the last digit of the answer.

**Example 3.8**

|       | 4             | 9            | 7          |
|-------|---------------|--------------|------------|
| 4-2=2 | (9-4)2+9=(1)9 | (9-9)2+7=7   | (10-7)2=6  |
| 2+(1) | 9             | 7            | 6          |
| 3     | 9             | 7            | 6          |

**Rule 3.8 (Multiplication by 4)** Subtract the right-hand figure of the given number from ten and add 5 if that digit is odd. Taking the remaining numbers up to the last 1, subtract it from 9, add half of its neighbour plus another 5 if the number is odd. Under the zero in front of the given number, write half the neighbour of this zero minus 1.

**Example 3.9**

|            | 4         | 9           | 7          |
|------------|-----------|-------------|------------|
| (4over2)-1=1 | 9-4+4=9 | 9-9+3+5=7   | 10-7+5=8   |
| 1          | 9         | 8           | 8          |
| 1          | 9         | 8           | 8          |

**Rule 3.9 (Multiplication by 3)** Subtract the right-hand figure of the given number from ten and double the result plus another 5 if the number is odd. Taking the remaining numbers up to the last 1, subtract it from 9 and double the result then add half of its neighbour plus another 5 is the number is odd. Finally, divide the left-hand figure by half then subtract 2. That gives us the final digit of the answer.

**Example 3.10**

|           | 4            | 9           | 7             |
|-----------|--------------|-------------|---------------|
| (4over)-2=0 | (9-4)2+4+5=(1)4 | (9-9)2+3+5=8 | (10-7)2+5=(1)1 |
| 0+(1)     | 4            | 8+(1)       | 1             |
| 1         | 4            | 9           | 1             |

For more comprehensive information on the Trachtenberg system, as well as the algorithms of multiplication of any digit numbers, the reader is referred to Trachtenberg [22].

**4. Conclusions and future research**

We have discussed the possibilities of using the Trachtenberg rapid mental computation system as a tool for improving the algorithmic thinking of elementary school students. It is well-known that one of the problems of the computer science propaedeutic course is developing algorithmic and logical thinking. In our opinion, algorithmic thinking development can be started in elementary school in mathematics lessons, since to think algorithmically means the ability to solve problems of various origins, requiring an action plan to achieve the desired result, and only mathematics can be considered as the universal language for describing the laws of nature.

Our implementation of several multiplication algorithms in computer algebra system Matlab 2011a has shown that the Trachtenberg rapid mental computation system can decrease the computation cost of multiplying numbers. So, there might be hope that computations in computer algebra systems based on the Trachtenberg system can accelerate operations with numbers. All these issues would offer considerable scope for our future work.

**References**
1. Agrawala, V. S. (1992). Vedic mathematics. Motilal Banarsidass Publ., India.
2. Benjamin, A., Shermer, M. (2006). Secrets of Mental Math: The Mathematicians Guide to Lightning Calculation and Amazing Math Tricks. California, USA: Three Rivers Press.
3. Bill Handley. (2007). Speed Math for Kids: The Fast, Fun Way To Do Basic Calculations, California, USA: John Wiley & Sons.
4. Bill Handley. (2012). Speed Mathematics, California, USA: John Wiley & Sons.
5. Burton, B. A. (2010). Encouraging algorithmic thinking without a computer. Olympiads in Informatics 4, 3 – 14.
6. Cha, S. E., Jun, S. J., Kwon, D. Y., Kim, H. S., Kim, S. B., Kim, J. M., Kim, Y. A., Han, S. G., Seo, S. S., Jun, W. C., Kim, H. C., Lee, W. G. (2011). Measuring achievement of ICT competency for students in Korea. Computers and Education 56 (4), 990 – 1002.
7. Chuck Lotta. (2000). Fast & Fun Mental Math: 250 Quick Quizzes to Sharpen Math Skills Every Day of the School Year. New-York: Scholastic Inc.

8. Cooper, S., Dann, W., Pausch, R. (2000). Developing algorithmic thinking with Alice. In: Information System Educators Conference. pp. 506 – 539.
9. Ewa Kolczyk. (2008). Algorithm - Fundamental Concept in Preparing Informatics Teachers. In Mittermeir, Roland and Syslo, Maciej (Eds.), Informatics Education - Supporting Computational Thinking in Lecture Notes in Computer Science, pages 265 - 271. Springer Berlin / Heidelberg.
10. Futschek, G., (2006). Algorithmic thinking: The key for understanding computer science. In: Mittermeir, R. (Ed.), Informatics Education The Bridge between Using and Understanding Computers. Vol. 4226 of Lecture Notes in Computer Science. Springer Berlin / Heidelberg, pp. 159 – 168.
11. G. Futschek, J. Moschitz. (2011). Learning Algorithmic Thinking with Tangible Objects Eases Transition to Computer Programming. Informatics in Schools, Contributing to 21st Century Education, pages 155 - 164, Springer-Verlag Berlin Heidelberg.
12. Harry Lorayne. (1992). Miracle Math: How to Develop a Calculator In Your. New York: Barnes & Noble.
13. Inge Schwank. (1993). On the Analysis of Cognitive Structures in Algorithmic Thinking. Journal of Mathematical Behavior, 12(2), 209 – 231.
14. Jakow Trachtenberg. (1960). The Trachtenberg Speed System of Basic Mathematics. Garden City, New York: Doubleday and Company, Inc.
15. Julius, E. H. (1992). How & Tips, 30 Days to Number Power. California, USA: John Wiley & Sons.
16. Knuth, Donald E. (1985). Algorithmic Thinking and Mathematical Thinking. The American Mathematical Monthly, 92(3), 170 – 181.
17. Milkova, E. (2005). Developing of algorithmic thinking: the base of programming. International Journal of Continuing Engineering Education and Life Long Learning 15 (3-6), 135 – 147.
18. Pickover, C. A. (2002). The Mathematics of Oz: Mental Gymnastics from Beyond the Edge. United Kingdom: Cambridge University Press.
19. Schwank, I. (1993). On the analysis of cognitive structures in algorithmic thinking. Journal of Mathematical Behavior 12 (2), 209 – 231.
20. Stigler, J. W. (1984). "Mental abacus": The effect of abacus training on Chinese children's mental calculation. Cognitive Psychology 16 (2), 145.
21. Štěpán Hubálovský, Eva Milková, Pavel Pražák. (2010). Modeling of a real situation as a method of the algorithmic thinking development and recursively given sequences. WSEAS Transactions on Information Science and Applications, 7(8), 1090 - 1100.
22. Trachtenberg, J. (1960). The Trachtenberg Speed System of Basic Mathematics. Doubleday and Company, Inc., Garden City, NY, USA.

**Быстрая интеллектуальная вычислительная система как инструмент развития алгоритмического мышления учеников начальной школы**

[1] Рушан Зиатдинов
[2] Саджид Мусса

[1-2] Университет Фатих, Турция
34500 Стамбул, ул. Бюуюкшекмеке
[1] Кандидат наук по математическому моделированию, доцент
E-mail: rushanziatdinov@gmail.com
[2] Аспирант

**Аннотация.** В данной работе описываются возможности использования быстрой интеллектуальной вычислительной системы в начальном образовании. Система состоит из ряда готовых операций, позволяющих каждому быстро произвести математические вычисления. Эти операции являются, по сути, простыми алгоритмами, которые могут развить или улучшить мышление учеников. Использование быстрой интеллектуальной вычислительной системы позволяет создать основу для обучения информатике в средней школе.

**Ключевые слова:** алгоритмическое мышление; быстрое интеллектуальное вычисление; система Трахтенберга; информатика; математика.